\documentclass[aps, prper, preprint, superscriptaddress]{revtex4-2}

\usepackage{graphicx} 
\usepackage{subcaption}
\usepackage{amsmath}  
\usepackage{hyperref} 
\usepackage{booktabs}
\usepackage{xcolor}   

\begin{document}

\title{\textit{PhysTrackX}: Open-Source Software for Kinematics Tracking for the Physics Laboratory}
\author{Isam Abdullah Balghari}
\affiliation{Department of Physics, Syed Babar Ali School of Science and Engineering, Lahore University of Management Sciences (LUMS), Opposite sector U, D.H.A, Lahore 54792, Pakistan}

\author{Muhammad Sabieh Anwar}\email{To whom correspondence should be addressed: sabieh@lums.edu.pk}
\affiliation{Department of Physics, Syed Babar Ali School of Science and Engineering, Lahore University of Management Sciences (LUMS), Opposite sector U, D.H.A, Lahore 54792, Pakistan}
\affiliation{Lahore Science Foundry, Khwarizmi Science Society, Pakistan}
\date{\today}

\begin{abstract}
Video tracking is a staple of the undergraduate physics laboratory, enabling students to extract quantitative kinematic data from real-world phenomena. In this article, we introduce \textit{PhysTrackX}, an open-source video analysis framework designed to reduce manual tracking overhead and streamline multi-modal data acquisition. Built with a modular graphical interface, \textit{PhysTrackX} integrates video trimming, spatial calibration, and coordinate system definition with automated, machine-learning-based object tracking. It introduces an optical character recognition (OCR) module for synchronizing kinematic trajectories with external digital sensor displays, an interactive geometry plugin for in-situ spatial measurements, and integrated plotting and data export tools. We outline the software architecture, benchmark its practical advantages and demonstrate its capabilities through three mechanics and thermodynamics case studies, thus offering a useful complement to the Open Source Physics (OPS) based Tracker. Pre-compiled standalone executables and source code are freely provided to ensure immediate classroom deployment without external runtime dependencies.
\end{abstract}

\maketitle

\section{Introduction}
Instructional physics laboratories bridge abstract mathematical models and physical reality by using video analysis to extract kinematic data \cite{brown2008tracker, wee2015video}. While general-purpose tools like ImageJ excel in static biomedical imaging and Kinovea \cite{kinovea} support biomechanical annotation, the Open Source Physics (OSP) Tracker \cite{brown2008tracker} serves as \emph{the} primary benchmark for quantitative 2D physical modeling.
Most of these software rely on manual point marking or classical template matching algorithms that degrade under severe motion blur, rapid acceleration, local illumination changes, or occlusions. Furthermore, conventional tools lack native mechanisms to align multi-modal experimental data—such as visual kinematics and standalone digital sensor readouts (e.g., multimeters or pressure gauges)—forcing tmanual synchronization and creating ``tool friction" that distracts students from core physical concepts \cite{van1991learning}.

To address some of these practical bottlenecks, we introduce \textit{PhysTrackX}, an open-source video tracking and kinematic analysis suite designed for instructional laboratories. The software offers three core contributions. First, it integrates a lightweight, anchor-free Siamese neural network (NanoTrack~\cite{nanotrack}) executed via the Open Neural Network Exchange (ONNX) runtime~\cite{onnx}, delivering drift-resistant multi-target tracking on standard laptop CPUs without requiring GPU acceleration (see Sec.~S1 of the Supplemental Material~\cite{suppmat} for more technical details). Second, it incorporates an optical character recognition (OCR) pipeline to read in-frame digital displays frame-by-frame, directly synchronizing external sensor streams with spatial kinematics. Third, it provides a unified workflow encompassing proxy video generation for responsive frame scrubbing, interactive spatial calibration, and automated kinematic plotting. A step-by-step operational manual is provided in Sec.~S2 of the Supplementary Material~\cite{suppmat}. The software is also readily downloadable from our website~\cite{physlab_website} which also provides a video tutorial.

In this article, no claim is made for \emph{superiority} over Tracker or for that matter, any other tool employed in the instructional laboratory. In fact, the authors use Tracker for many of their video tracking targets. We merely present PhysTrackX as an alternative open-source complementary software that utilizes modern libraries for robust tracking and which is built around a simple-to-use workflow. For example, while Tracker uses RGB pixel differences across candidate patches~\cite{tracker_autotracker_help}, PhysTrackX targets objects by structural geometry and semantic contours via deep feature maps~\cite{siamese, nanotrack}. This feature based attraction has the potential to display robustness against losing track in the presence of shadows, glare and clutter~\cite{smeulders2014visual}. Another important feature of PhysTrackX is extracting numerical data from front panel displays of digital meters in a time-synchronous fashion.

\section{Core Software Features}

\begin{figure*}[t]
  \centering
  \includegraphics[width=\linewidth]{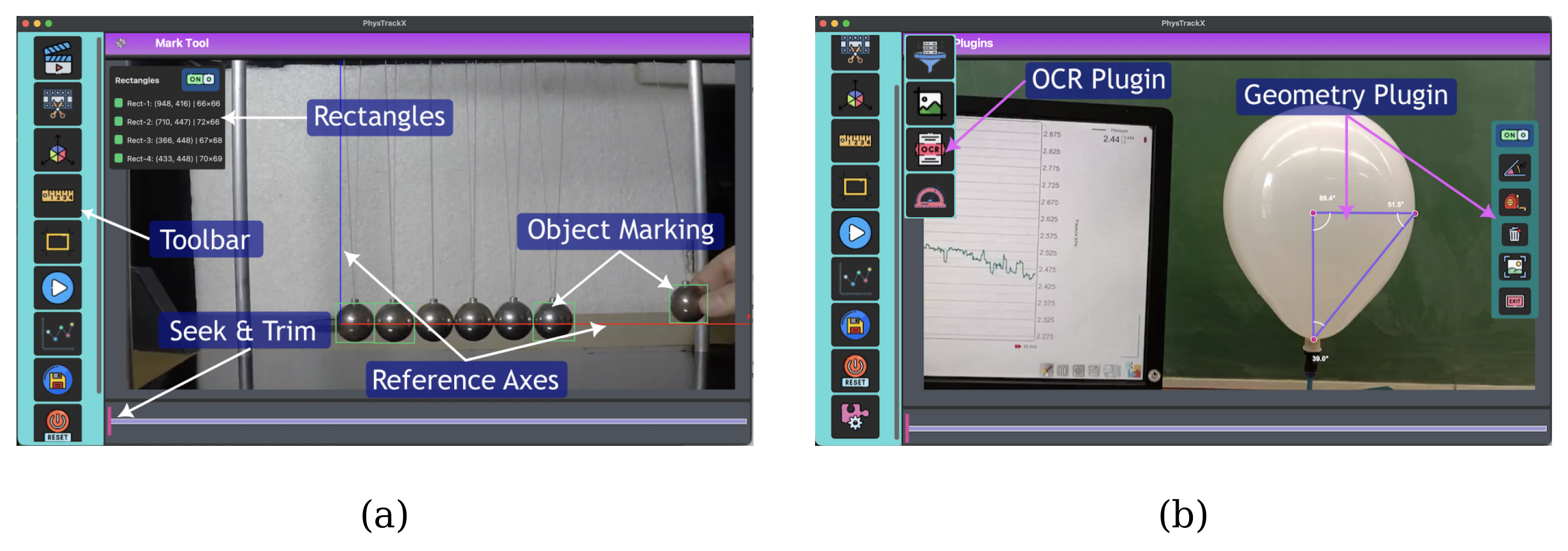}
  \caption{Overall workflow of \textit{PhysTrackX}. Figures show an overview of important features: (a) demonstrates the loading, trimming, object marking and calibration tools while (b) shows the OCR and geometry plugins.}
  \label{fig:software}
\end{figure*}

\textit{PhysTrackX} is developed in Python, leveraging the \texttt{customtkinter} \cite{customtkinter} library to render a streamlined graphical user interface (GUI) designed to fast track data acquisition process during laboratory exercises. The interface is described shortly. The video processing engine is powered by OpenCV \cite{opencv_library} and FFmpeg \cite{ffmpeg} which ensure smooth, lag-free frame-by-frame scrubbing, even when processing high-resolution footage captured by modern smartphones. A primary pedagogical utility of the software is anchored by its streamlined, modular toolset. Fig.~\ref{fig:software} shows an overview of the various modules in action. We now outline some practical features of the software while relegating more technical ``under-the-hood'' specifics to the Supplemental Material~\cite{suppmat}.

\subsection{Streamlined User Interface}
An important feature of \textit{PhysTrackX} is its streamlined GUI and conceptual workflow which we believe minimizes the learning curve for new users, carrying a natural progression of any video tracking pipeline. It starts with a built-in video loader and viewer alongside a video trimming tool that allows students to isolate a specific sub-video segment. It then provides tools to define the axis and physical scale, offers tracking and plotting tools that easily view the extracted data, followed by a save tool. It also features plugins that offer options to extract any numerical text data via the OCR tool, filters to aid in maintaining tracking stability in sub-optimal conditions—such as motion blur or poor laboratory lighting. These filters include Canny edge detection \cite{canny}, Gaussian blurring \cite{gaussianblur}, median filtering \cite{medianfilter}, bilateral filtering \cite{bilateralfilter}, and contrast and brightness enhancements directly to the video prior to tracking, all aimed to offer possibilities for high-fidelity data extraction. 

\subsection{Automated Object Tracking}
\textit{PhysTrackX} implements a lightweight Siamese neural network tracker~\cite{nanotrack, siamese} for automated multi-target tracking. Using the \texttt{Mark} tool, students draw bounding boxes around one or more regions of interest on the initial frame, after which the software autonomously tracks each target across the trimmed video sequence.

\subsection{Synchronized Optical Character Recognition}
A frequent operational challenge in laboratory video analysis is synchronizing spatial trajectories with readings from independent digital instruments, such as multimeters, electronic balances, or digital manometers. \textit{PhysTrackX} integrates an optical character recognition (OCR) plugin powered by the Tesseract engine~\cite{tesseract}. Analogous to object marking, students draw a stationary bounding box over any digital display visible within the video frame, as annotated in Fig.~\ref{fig:software}(b). During tracking, the software parses the numeric characters on each frame and appends the readings directly to the exported kinematic data table under synchronized timestamps.

\subsection{Geometric Visualization}
Visualizing coordinate axes, spatial scaling, and apparatus geometry aids experimental setup verification. The \textit{PhysTrackX} \texttt{Geometry} plugin enables students to overlay interactive line segments and triangles directly onto the video canvas. Using the calibrated pixel-to-length scale factor, the software calculates physical side lengths ($a, b, c$) and evaluates interior vertex angles via the law of cosines. This allows students to perform in-situ geometric measurements such as verifying ramp inclination angles or maximum pendulum angular displacements without physical protractors. An example is shown in Fig.~\ref{fig:software}(b).

\subsection{Data Management}
Correlating physical motion with real-time graphical representations reinforces kinematics graph interpretation~\cite{van1991learning, beichner1996impact}. To support immediate preliminary inspection, \textit{PhysTrackX} provides integrated \texttt{Plot} and \texttt{Save} utilities. The \texttt{Plot} tool generates time-series graphs of position, numerical velocity, and acceleration, as well as two-dimensional parametric trajectories ($x$ vs.\ $y$). The \texttt{Save} tool exports the calibrated spatial coordinates, synchronized OCR sensor readings, and timestamps to a comma-separated values (\texttt{.csv}) file for curve fitting and uncertainty analysis in any external software of the user's choice.

\section{Examples}
To demonstrate the practical and pedagogical efficacy of \textit{PhysTrackX}, we present three well-known physics laboratory implementations. The first two focus on the software's streamlined intuitive data extraction workflow along with the plotting tool, while the last features the digital sensor reading synchronization using the OCR plugin.

\subsection{Momentum Conservation and Rotational Transfer}

\begin{figure}[t!]
\centering
\includegraphics[width=\columnwidth]{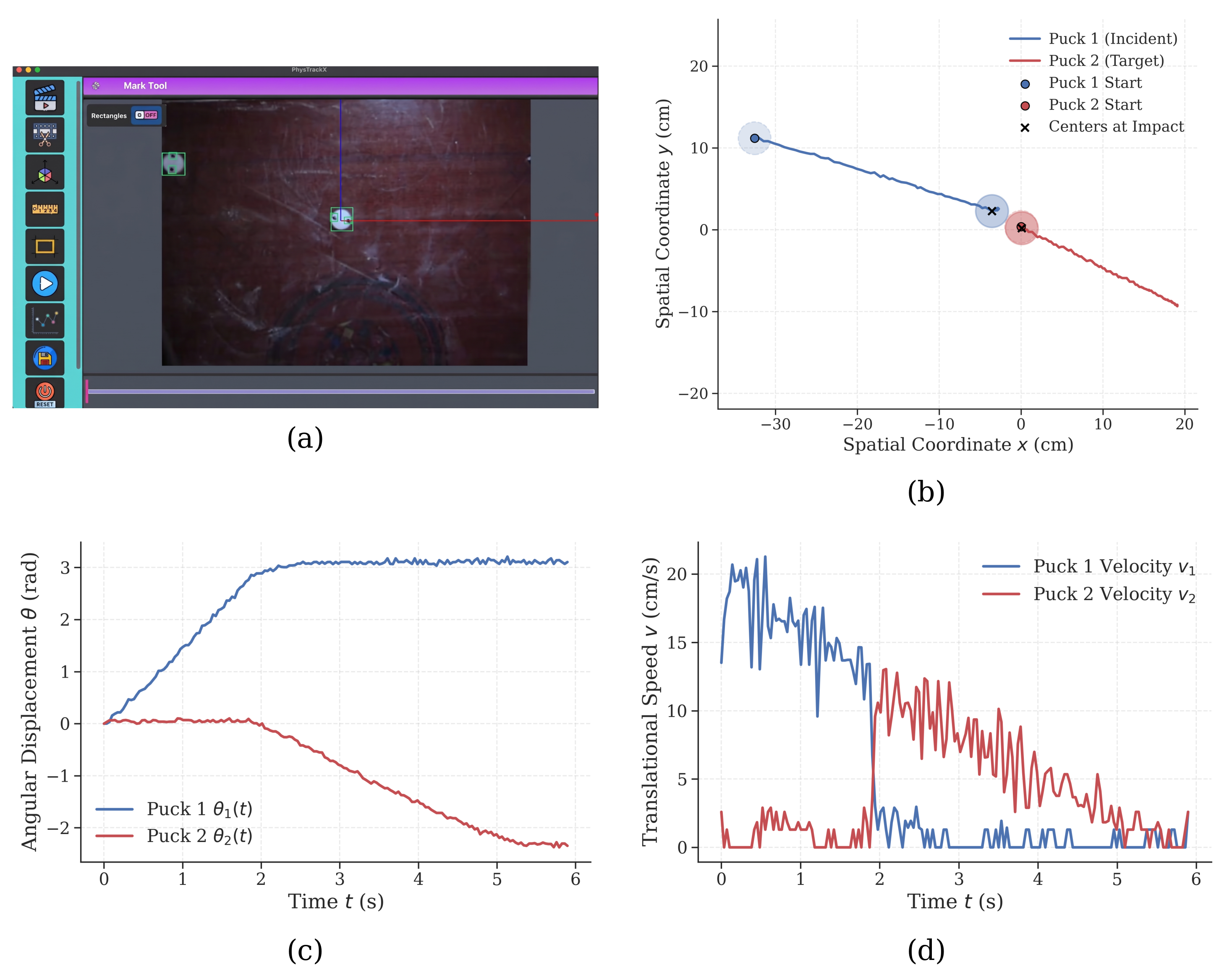}
\caption{Kinematic analysis of a two-body glancing carom puck collision using \textit{PhysTrackX}. (a) Overhead camera perspective showing the experimental carom board setup and target initialization. (b) Reconstructed two-dimensional spatial trajectories ($x$ vs $y$), where circular patches indicate puck dimensions at the initial release and at the instant of impact. (c) Unwrapped angular displacement $\theta(t)$ showing rotational transfer upon collision. (d) Translational speed profiles $v(t)$ of the incident (Puck 1) and target (Puck 2) pucks before and after impact.}
\label{fig:rotation_transfer}
\end{figure}

Collisions between two bodies in a plane constitute a foundational mechanics experiment in undergraduate physics laboratories~\cite{aji2022,williamson2000}. Here, we demonstrate the utility of \textit{PhysTrackX} in simultaneously extracting translational trajectories, linear speeds, and planar rotational kinematics from an inelastic glancing collision between two identical carom pucks. Each puck is marked with two high-contrast fiducial points for orientation tracking and placed on a level carom board beneath an overhead camera. The user interface with the video loaded in is depicted in Fig.~\ref{fig:rotation_transfer}(a).

The analysis workflow in \textit{PhysTrackX} begins by importing and temporally trimming the video clip to isolate the $6\text{ s}$ collision window. A Cartesian coordinate system is placed near the board center, and a pixel-to-centimeter calibration factor is established. Six bounding boxes are then initialized: two encompassing the puck bodies for center-of-mass tracking, and four isolating the individual surface fiducial markers to resolve planar orientation angles $\theta_1(t)$ and $\theta_2(t)$. Executing the tracking routine processes all six targets simultaneously across the sequence without manual frame-by-frame intervention, exporting time-resolved coordinates and timestamps to a structured CSV file.

The reconstructed planar trajectories [Fig.~\ref{fig:rotation_transfer}(b)] and translational speed profiles [Fig.~\ref{fig:rotation_transfer}(d)] capture the glancing collision at $t \approx 1.8\text{ s}$. Notably, the unwrapped angular displacement curves [Fig.~\ref{fig:rotation_transfer}(c)] clearly resolve spin transfer: prior to impact ($t < 1.8\text{ s}$), the incident striker (Puck~1) spins at a nearly constant angular velocity ($\omega_1 > 0$) while the target (Puck~2) remains at rest ($\theta_2 = 0$). Upon eccentric contact, tangential rim friction arrests the rotation of Puck~1 ($\theta_1 \approx \text{const}$ for $t > 2.0\text{ s}$) and imparts an opposite spin ($\omega_2 < 0$) to Puck~2. Furthermore, Fig.~\ref{fig:rotation_transfer}(d) illustrates two instructive experimental features for undergraduate laboratories: (i) raw finite-difference numerical differentiation ($v = |\mathrm{d}\mathbf{r}/\mathrm{d}t|$) amplifies frame-to-frame pixel quantization noise, motivating students to extract pre- and post-impact velocities by fitting linear slopes to position data immediately adjacent to $t \approx 1.8\text{ s}$, and (ii) the steady post-impact deceleration of both pucks to rest ($t \in [2.0, 5.5]\text{ s}$) allows students to quantify dry sliding friction on the board while verifying momentum conservation across the brief collision impulse.

\subsection{Galileo’s Damped Oscillator}

\begin{figure*}[t]
\centering
\includegraphics[width=\linewidth]{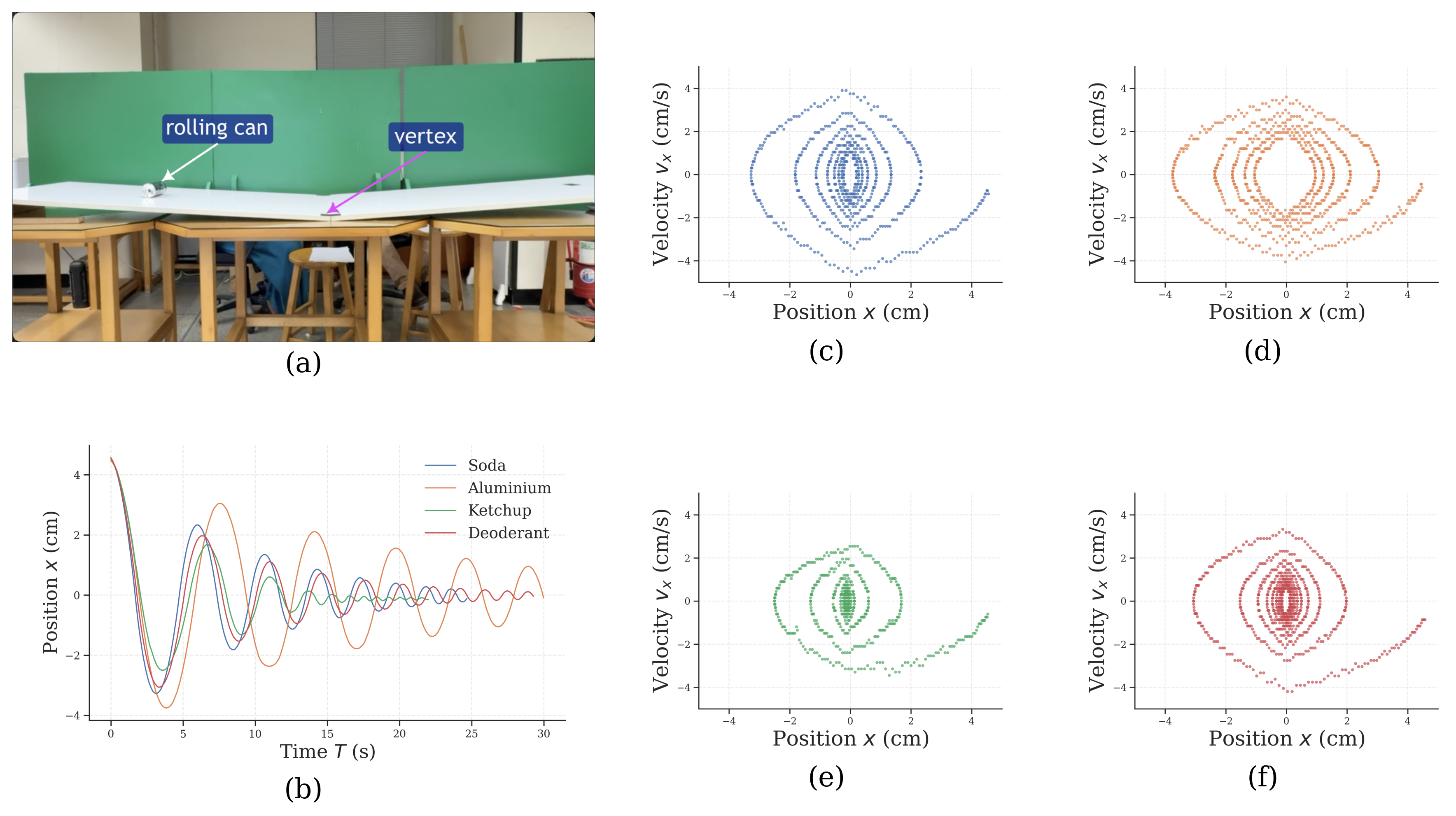}
\caption{Experimental apparatus and kinematic tracking data for Galileo's damped oscillator. (a) Shows the physical experimental setup, (b) combines position-time trajectories across four compositions, and (c--f) show the phase-space maps ($v$ vs.\ $x$) corresponding to soda, aluminum, ketchup and deodorant illustrating fluid shear-induced dissipation.}
\label{fig:galileo}
\end{figure*}

We next examine ``Galileo's Damped Oscillator,'' a classic mechanics experiment investigating how internal fluid viscosity within cylindrical shells rolling along a V-shaped track governs mechanical energy dissipation~\cite{galileo}. The track is constructed from two inclined planes joined at a shallow vertex [Fig.~\ref{fig:galileo}(a)]. Four cylinders of differing internal composition---a solid aluminum rod and three hollow containers filled with soda, ketchup, and deodorant---are released from rest and recorded with a tripod-mounted smartphone camera.

After calibrating the coordinate origin at the vertex and scaling the workspace in \textit{PhysTrackX}, each cylinder is tracked across multiple oscillation periods. Figure~\ref{fig:galileo}(b) overlays the resulting position-time curves $x(t)$, while Figs.~\ref{fig:galileo}(c)--(f) display the corresponding phase-space trajectories ($v_x$ vs.\ $x$) for soda, solid aluminum, ketchup, and deodorant, respectively. The phase-space portraits visually distinguish the dissipative regimes: the rigid aluminum cylinder exhibits a tightly wound spiral corresponding to weak rolling resistance, whereas the fluid-filled cylinders exhibit rapid inward spiraling driven by internal fluid sloshing and viscous shear dissipation.

\subsection{Double Balloon Experiment}

\begin{figure*}[h]
\centering
\includegraphics[width=\linewidth]{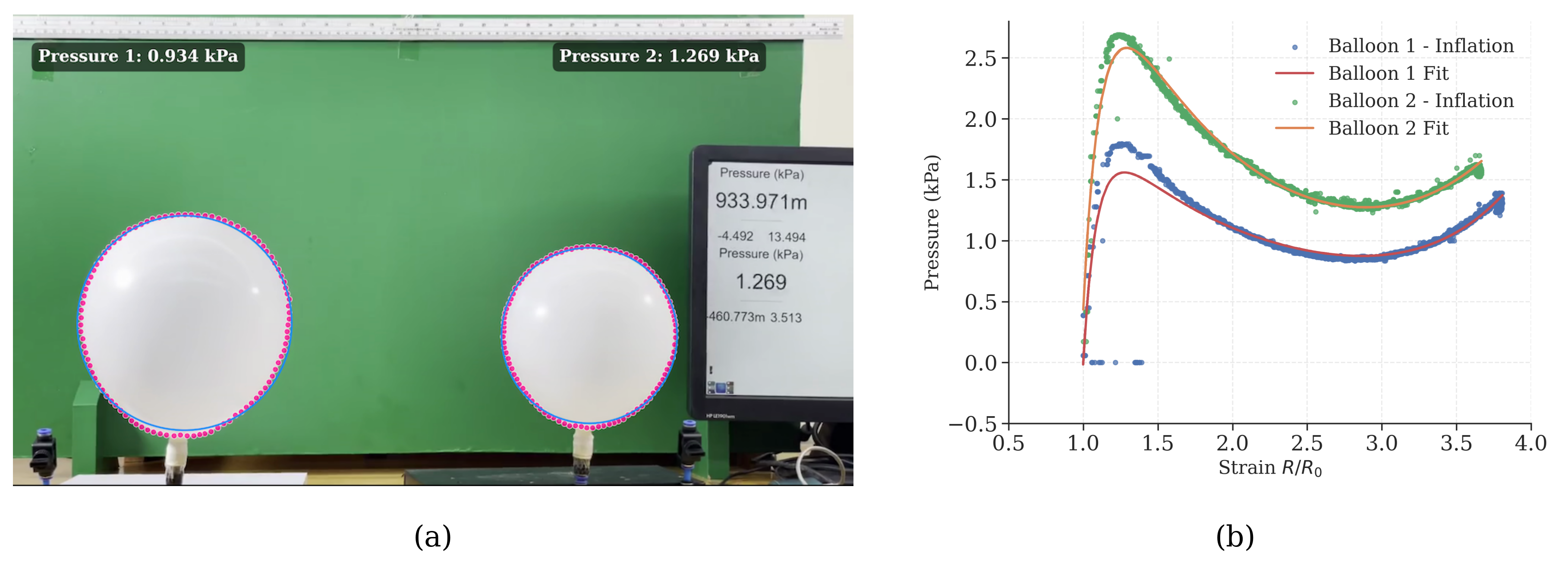} 
\caption{Demonstration of OCR integration. (a) Interface showing the simultaneous tracking of balloon boundary and sensor readings. (b) Plots of extracted radiuses vs pressure values.}
\label{fig:ocr_integration}
\end{figure*} 

To demonstrate multi-modal telemetry synchronization, we apply the integrated OCR module to the classic ``Two-Balloon Experiment''~\cite{levin2004two}, in which two interconnected rubber balloons of unequal initial radiuses exchange air through a connecting valve. Characterizing the non-linear hyperelastic constitutive behavior of the rubber membrane requires simultaneous measurement of the balloon radius $R(t)$ and the internal gauge pressure $P(t)$ displayed on an in-frame digital manometer~\cite{merritt1978pressure}.

As shown in Fig.~\ref{fig:ocr_integration}(a), \textit{PhysTrackX} concurrently tracks the expanding circular boundaries of both balloons while extracting the live digital pressure readout in each frame via the OCR plugin. Synchronizing the spatial and manometer data streams from a single video file enables direct construction of the pressure--stretch curve [Fig.~\ref{fig:ocr_integration}(b)], where the radial stretch ratio $R/R_0$ is fitted to the Mooney--Rivlin hyperelastic membrane model~\cite{mooney1940theory}. The automated trajectory accurately captures the steep initial Hookean rise, the local pressure maximum, intermediate softening, and the high-stretch strain-hardening regime.

\section{Conclusion and Availability}
\textit{PhysTrackX} provides an accessible, open-source environment for video kinematics and synchronized digital telemetry in undergraduate physics laboratories. By combining a CPU-optimized Siamese neural network tracker with an integrated OCR engine and an intuitive calibration workflow, the software reduces manual frame-by-frame clicking and eliminates tedious hand-transcription of external digital displays. These capabilities allow students to spend less of their precious lab time troubleshooting tracking failures and more in capturing and modeling physical phenomena, evaluating experimental uncertainties, and testing physical principles.

In alignment with open-science practices for physics education, the source code for \textit{PhysTrackX} is publicly available and distributed under the Business Source License (BSL), making it completely free for academic, instructional, and non-commercial research use. Pre-compiled standalone Windows executables (\texttt{.exe}), the Python source code, and compilation instructions are hosted on GitHub~\cite{github-isam,physlab_website}.

\section*{Author Contributions}
I.A.B wrote the code whereas I.A.B and M.S.A worked together on the conceptual design of the software, performed the experiments, analzyed data, wrote the manuscript, tested and debugged the software.

\section*{Acknwoledgments}
The authors like to thank Shirin Abbas for help in preparing the video tutorial and the students who helped in data generation with the software. Tayyab Iftikhar also demonstrated the software in workshops organized by the Khwarizmi Science Society held in Jhelum and Islamabad.

\bibliography{anwarphystrackx}

\end{document}